# Multiclass Spinal Cord Tumor Segmentation on MRI with Deep Learning

**Authors**

Andreanne Lemay[1,2], Charley Gros[1,2], Zhizheng Zhuo[3], Jie Zhang[3], Yunyun Duan[3], Julien Cohen-Adad[1,2,4], Yaou Liu[3]

**Affiliations**

*1 - NeuroPoly Lab, Institute of Biomedical Engineering, Polytechnique Montreal, Montreal, QC, Canada*
*2 - Mila, Quebec AI Institute, Canada*
*3 - Beijing Tiantan Hospital, Capital Medical University, Beijing, China*
*4 - Functional Neuroimaging Unit, CRIUGM, Université de Montréal, Montreal, QC, Canada*

**CORRESPONDING AUTHORS**
Yaou Liu M.D; PhD
Executive Director of Department of Radiology,
Beijing Tiantan Hospital, Capital Medical University
No.119 Southern 4th Ring Road West,
Fengtai District, Beijing 100070, China
e-mail: yaouliu80@163.com; liuyao@bjtth.org
Phone/Fax: +86 10 59975358

Julien Cohen-Adad
Dept. Genie Electrique, L5610
Ecole Polytechnique
2900 Edouard-Montpetit Bld
Montreal, QC, H3T 1J4, Canada
e-mail: jcohen@polymtl.ca
Phone: 514 340 5121 (office: 2264);  Skype: jcohenadad

**Abbreviations**
CNN: convolutional neural network
IMSCT: intramedullary spinal cord tumor
T1w: T1-weighted
T2w: T2-weighted

# Abstract

Spinal cord tumors lead to neurological morbidity and mortality. Being able to obtain morphometric quantification (size, location, growth rate) of the tumor, edema, and cavity can result in improved monitoring and treatment planning. Such quantification requires the segmentation of these structures into three separate classes. However, manual segmentation of 3-dimensional structures is time-consuming and tedious, motivating the development of automated methods. Here, we tailor a model adapted to the spinal cord tumor segmentation task. Data were obtained from 343 patients using gadolinium-enhanced T1-weighted and T2-weighted MRI scans with cervical, thoracic, and/or lumbar coverage. The dataset includes the three most common intramedullary spinal cord tumor types: astrocytomas, ependymomas, and hemangioblastomas. The proposed approach is a cascaded architecture with U-Net-based models that segments tumors in a two-stage process: locate and label. The model first finds the spinal cord and generates bounding box coordinates. The images are cropped according to this output, leading to a reduced field of view, which mitigates class imbalance. The tumor is then segmented. The segmentation of the tumor, cavity, and edema (as a single class) reached $76.7 \pm 1.5\%$ of Dice score and the segmentation of tumors alone reached $61.8 \pm 4.0\%$ Dice score. The true positive detection rate was above 87% for tumor, edema, and cavity. To the best of our knowledge, this is the first fully automatic deep learning model for spinal cord tumor segmentation. The multiclass segmentation pipeline is available in the Spinal Cord Toolbox (https://spinalcordtoolbox.com/). It can be run with custom data on a regular computer within seconds.

# Keywords

# 1. Introduction

Intramedullary spinal cord tumors (IMSCT) represent 2 to 5% of all central nervous system tumors (M Das et al., 2020). This relatively low prevalence contributes to the difficulty in understanding this malignant pathology (Claus et al., 2010). The first step to a better understanding of the disease is an improved characterization of the tumor. Segmentation informs the healthcare specialists on the tumor's position, size, and growth rate leading to quantitative monitoring of the tumor's progression. In addition, characterizing the edema and cavity (i.e., syrinx) associated with the tumor is clinically-relevant (Balériaux, 1999; Kim et al., 2014; M Das et al., 2020). Manual labeling is tedious for clinicians and prone to intra- and inter-rater variability. Fully automatic segmentation models overcome these issues. Although automatic methods to segment brain tumors are numerous, there is currently, to the best of our knowledge, no automatic model to segment IMSCT. Heterogeneity in tumor size, intensity, location, in addition to images varying in resolution, dimensions, and fields of view represent a challenge for the segmentation task.

## 1.1. Heterogeneity in spinal cord tumor characteristics

This study covers the three main tumor types (95% of IMSCT (Balériaux, 1999)): astrocytoma, ependymoma, and hemangioblastoma, with cervical, thoracic, and lumbar coverage (Figure 1). On T2-weighted (T2w) scans, all tumor types display isointense to hyperintense signals (Figure 1A), and on T1w-weighted (T1w) scans from isointense to hypointense signals, with well or ill-defined boundaries (Baker et al., 2000; Kim et al., 2014). In the case of isointense tumors or ill-defined margins, it is challenging to segment the lesion properly. Gadolinium-enhanced T1w MRI yields a hyperintense signal from the tumor, but the enhancement patterns are different depending on the tumor type. For example, astrocytomas usually present partial, moderate (Figure 1E), or no enhancement (Figure 1D) (Balériaux, 1999), while hemangioblastomas generally yield intense enhancement (Figure 1F) (Baker et al., 2000). All tumor types do not display the same characteristic on MRI scans which is a challenge when developing a robust deep learning model. Tumors can be associated with tumoral or non-tumoral cystic components (i.e., liquid-filled cavities in the spinal cord) (Figure 1 C and F) or extensive edema, but these cannot be systematically seen with the tumor (Baker et al., 2000; Balériaux, 1999; Kim et al., 2014). Finally, tumors can be present as a single (Figure 1A, 1B, 1D, 1E) or multiple lesions (Figure 1C and 1F) (Balériaux, 1999) and measure from less than 10 mm (Chu et al., 2001) up to 19 vertebral bodies long (Balériaux, 1999). This heterogeneity in IMSCT can be an obstacle to precisely delineate each component related to the tumor.

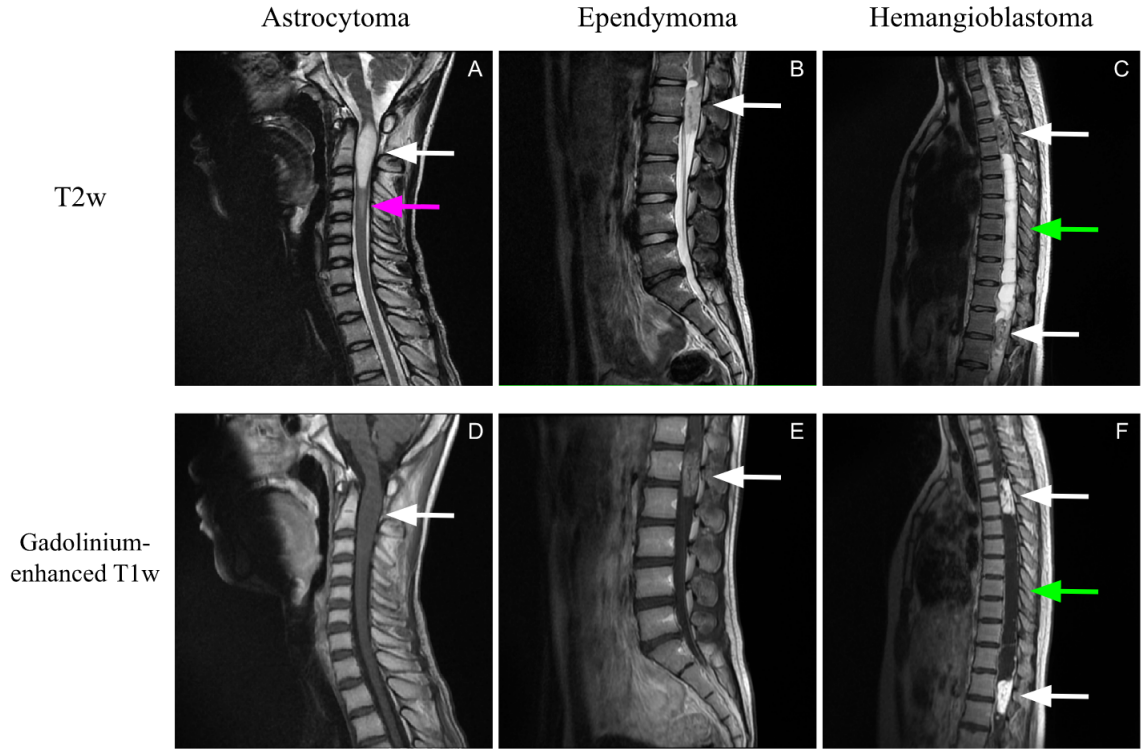

Figure 1: IMSCT heterogeneity in intensities, size, and longitudinal location. The first row (A, B, C) presents T2w scans, while the second row (D, E, F) are the gadolinium-enhanced T1w scans. The first column (A, D) displays an example of astrocytoma, the second column (B, E) an example of ependymoma, and the last column (C, F) an example of hemangioblastoma. The white arrows point to solid tumor components, the green arrows point to a liquid-filled cavity, and the pink arrow to edema.

## 1.2. Previous work

### 1.2.1. Brain tumor models

The annual BraTS challenge contributes to the progress of deep learning models for brain tumor segmentation. The challenge provides multi-contrast MRI scans (T1w, gadolinium-enhanced T1w, T2w, and FLAIR) of the brain with multi-label segmentations. Peritumoral edema, gadolinium-enhanced tumor, and non-enhancing and necrotic tumor core are identified and labeled by the deep learning models. The top-performing models on BraTS are mostly convolutional neural networks (CNNs) (Havaei et al., 2017; Isensee et al., 2019, 2018; Kamboj et al., 2018; Naceur et al., 2018). Three main types of CNN architecture are seen for brain tumor segmentation: patch-wise, semantic-wise, and cascaded (Akkus et al., 2017). The patch-wise architecture separates the image into subvolumes, where every patch predicts the class of the central pixel. Semantic-wise acts as an autoencoder (Isensee et al., 2019). The data is encoded to learn the more important features to class the pixels. The data is then decoded, resulting in a prediction for every pixel encoded. The advantage of semantic-wise compared to patch-wise is the training and inference speed since all pixels are predicted at once (Akkus et al., 2017). The last architecture family is a combination of CNN models where the output of a model becomes the input of the next.

Although they are both part of the central nervous system, brain tumors and spinal tumors need to be processed differently. A critical difference between the brain and the spinal cord is the size resulting in different fields of view. The spinal cord has an average diameter of 13.3 ± 2.2 mm at its largest point (transverse diameter in C5) (Frostell et al., 2016), while the length can range from 420 to 450 mm (Boonpirak and Apinhasmit, 1994). In contrast, the brain has more isotropic dimensions. This discrepancy leads to challenging decisions regarding cropping and patch sizes. In comparison with the brain, spinal cord imaging is also more hampered by respiratory and cardiac motion artifacts (Stroman et al., 2014). Another possible added difficulty of spinal tumor/edema/cavity segmentation is that, in comparison with the brain, these pathological presentations generally interfaces with the cerebrospinal fluid, making it sometimes difficult to separate them, especially on T2w scans, where the fluid appears as a hyperintense signal.

## 1.2.2. Spinal tumor models

Previous studies introduced models for segmenting tumors located in the spine (i.e., present in bone vs. spinal cord) (Hille et al., 2020; Reza et al., 2019; Wang, 2017). In 2020, Hille et al. developed a U-Net-based model to segment spine metastases (Hille et al., 2020). They reached a Dice score of 77.6% on average. Reza et al. 2019 presented a cascaded architecture for spine chordoma tumor segmentation (i.e., rare tumor type usually in the bone near the spinal cord and the skull base) (Reza et al., 2019). Their approach yielded a median absolute difference of 48%. Even if spine tumors and IMSCT are both present in the spine area, they exhibit different intensities and sizes and are juxtaposing with different tissue types (i.e., bone tissue vs. neuronal tissue), making spine segmentation models not applicable to the spinal cord. Also, previous models for spine tumors are single class (i.e., segmentation of tumor only), while IMSCT is associated with multiple components that should be segmented separately. Moreover, the trained models are not publicly available hence cannot directly benefit the medical community wanting to apply the model on new data.

## 1.2.3. Cascaded architecture in spinal cord tasks

State-of-the-art pipelines for medical imaging tasks exploited cascaded architectures (Akkus et al., 2017; Christ et al., 2017; Gros et al., 2019; Hussain et al., 2017). A rationale of cascaded models is to isolate the region of interest with a first CNN and then segment the desired structure. Gros et al. 2019 benefits from this approach for spinal cord multiple sclerosis segmentation (Gros et al., 2019). The first CNN finds the centerline while the second performs segmentation. This helped to limit class imbalance and focus the task on the spinal cord.

## 1.3. Study scope

In this work, we present a multi-contrast (gadolinium-enhanced T1w and T2w) segmentation model for IMSCT. We chose a two-stage cascaded architecture composed of two U-Nets (Ronneberger et al., 2015) (i.e., semantic-wise CNNs). The first step consists of localizing the region of interest, i.e., the spinal cord. The second step labels the structures of IMSCT. Both models are based on the modified 3D U-Net implemented by Isensee et al. for the 2017 BraTS challenge (Isensee et al., 2018). The framework segments the three main components associated with IMSCT: the enhanced and non-enhanced tumor component, liquid-filled cavities, and edema. The choice of a multi-contrast design was motivated by the relevance of both contrasts (gadolinium-enhanced T1w and T2w) for delineating all components of the IMSCT (Baker et al., 2000). In clinical routine, both contrasts are generally acquired (Balériaux, 1999). Our pipeline is robust to varying resolution, fields of view, and tumor sizes. The cascaded models were trained on the three common IMSCT types, which also happen to present very different image characteristics (shape, contrast, presence of edema/cavity). We integrated the model in the SCT open-source software (De Leener et al., 2017). Hence, the pipeline can be easily applied to custom data within seconds in a single command-line or via the graphical user interface (GUI) available in SCT (v5.0 and higher).

# 2. Material and methods

## 2.1. Dataset

The data used for this experiment includes 343 MRI scans acquired from Beijing Tiantan Hospital, Capital Medical University from October 2012 to September 2018, with heterogeneous vertebral coverage (cervical, thoracic, and lumbar), from patients diagnosed with spinal cord tumor: astrocytoma (101), ependymoma (122), and hemangioblastoma (120). T2-weighted (T2w) and Gadolinium-enhanced T1-weighted (T1w) images were available for each patient, as well as the manual segmentation for the tumor, edema, and cavity. Table 1 presents the demographic data. The native resolution of the sagittal scans was, on average, 0.60 mm x 0.60 mm x 3.68 mm.

Table 1: Demographic information of patients by tumor type. First row: number of subjects. Second row: sex distribution where M is male and F female. Third row: median age in years. Fourth row: Age range in years (minimum-maximum). The last column presents the demographic information for all tumor types combined.

|  | Astrocytoma | Ependymoma | Hemangioblastoma | All |
| --- | --- | --- | --- | --- |
| **Subjects** | 101 | 122 | 120 | 343 |

| Sex | 60M:41F | 70M:52F | 68M:52F | 198M:145F |
|---|---|---|---|---|
| **Median age** | 30 | 42 | 35.5 | 37 |
| **Age range (min-max)** | 1-63 | 8-76 | 8-66 | 1-76 |

## 2.2. Data preparation

A mask identifying the spinal cord was generated for the dataset with the SCT tools (De Leener et al., 2017). The centerline of the spinal cord for the images was manually identified. From the centerline, a mask of 30 mm diameter was automatically generated. This value was chosen in accordance with the average spinal cord diameter with an extra buffer to ensure full coverage of the spinal cord on the right-left and anterior-posterior axes. T1w images were registered (Avants et al., 2009) onto the T2w scans using affine transformations with the cross-correlation metric on the SCT software. Every registration has been manually verified and corrected if needed. Manual labels of vertebral discs were added and used to coregister the subjects when the first registration method failed.

## 2.3. Processing pipeline

Data preprocessing and model training was implemented with ivadomed (Gros et al., 2020b). Data is preprocessed before training or inference. The preprocessing steps are included in the model's pipeline. The resolution of the sagittal images are set to 1 mm (superior-inferior), 1 mm (anterior-posterior), 2 mm (right-left). This choice of resolution was based on preliminary investigations and is a compromise between computational time and segmentation precision. The resampled images were cropped voxel-wise with dimensions of 512 x 256 x 32, which corresponds to a bounding box of 51.2 cm x 25.6 cm x 6.4 cm applied at the center of the field of view. These dimensions are consistent with the adult spinal cord anatomy and allow for slight angulation in the right-left direction (e.g., in cases of scoliosis). In cases where the field of view is smaller in one or more axes, the image is zero-padded instead of cropped. The intensity of each scan was normalized by subtracting the average intensity over the standard deviation. The framework of the model is a cascaded architecture composed of two steps (Figure 2). The first step aims to localize the spinal cord and crop the image around the spinal cord mask with a 3D bounding box (Figure 2 step 1). The cropped image is used as input for the second step of the pipeline being the tumor segmentation task (Figure 2 step 2). The first step reduces the field of view on the tumor and makes the model robust for MRI with varying fields of view or dimensions. Smaller images also lead to faster training and inference in addition to mitigate class imbalance. For the tumor segmentation model, the cropping is done with the spinal cord mask. Thus, the cropping size is different for every patient according to the field of view, dimensions of the image, and size of the spinal cord. The segmentation yields 4 different labels: tumor core (enhanced and non-enhanced), cavity, edema, and the whole tumor composed of all structures together.

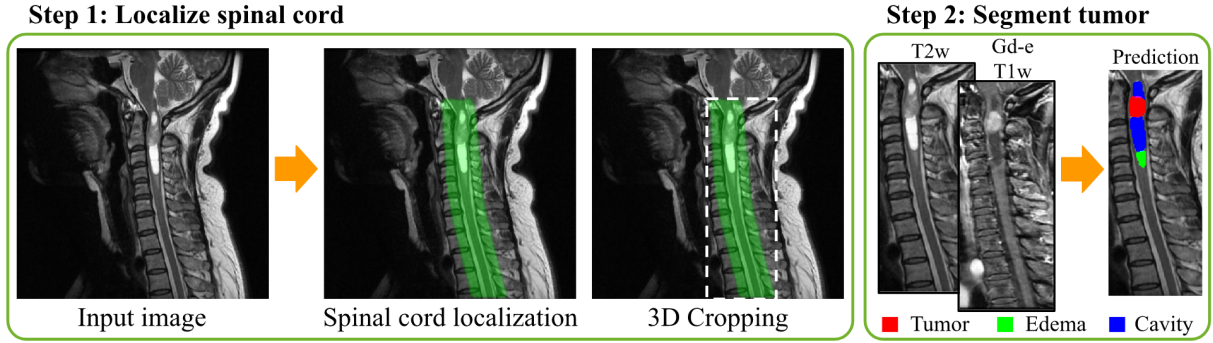

Figure 2: Fully automatic spinal cord tumor segmentation framework. Step 1: The spinal cord is localized using a 3D U-Net and the image is cropped around the generated mask. Step 2: The spinal cord tumors are segmented.

## 2.4. Model

A modified 3D U-Net (Isensee et al., 2018) architecture was used for both spinal cord localization and tumor segmentation. The main differences with the original 3D U-Net (Çiçek et al., 2016) reside in the addition of deep supervision modules (Kayalibay et al., 2017), the use of dropout layers (Srivastava et al., 2014), instance normalization (Ulyanov et al., 2016) (instead of batch normalization), and leaky ReLU (Maas et al., 2013) (instead of ReLU). Table 2 enumerates the main training parameters used for both models. Initial hyperparameter optimization was performed to find these values. Sigmoid was used as the final activation of the model, even for the multiclass model, to allow the prediction of a class composed of all structures.

Table 2: Training parameters for spinal cord localization and tumor segmentation model. Abbreviation: Gd-e: Gadolinium-enhanced.

| Parameters | Model 1: Spinal cord localization | Model 2: Tumor segmentation |
|---|---|---|
| Input | T2w | Gd-e T1w + T2w (multi-contrast) |
| Loss | Dice loss (Milletari et al., 2016) | Multiclass Dice loss |
| Batch size | 1 | 8 |
| Patch size | 512x256x32 | 128x128x32 |
| Stride | - | 64x64x32 |
| Early stopping | 50 | 50 |
| Learning rate | 0.001 | 0.001 |

| Learning rate scheduler | Cosine Annealing | Cosine Annealing |
|---|---|---|
| Depth | 4 | 4 |
| Number of base filters | 8 | 16 |
| **Data augmentation** | | |
| Rotation | ± 5 degrees | ± 5 degrees |
| Scaling | ± 10% | ± 10% |
| Translation | ± 3% | ± 3% |

## 2.4.1. Spinal cord localization

The spinal cord localization model was trained with a Dice loss (Milletari et al., 2016) using a single patch input of size 512x256x32 on T2w MRI scans. Preliminary experiments with a multi-contrast model did not improve the performance: a single channel model was hence used for this task. Due to the large size of each 3D patch, the batch size was limited to 1. The training lasted for a maximum of 200 epochs with an early stopping of 50 epochs with an epsilon of 0.001 (i.e., minimal loss function difference to be considered as an improvement). The initial learning rate was 0.001, and the cosine annealing scheduler was used. The number of downsampling layers (i.e., the depth) was set to 4, and the number of base filters (i.e., the number of filters used on the first convolution) to 8. Finally, random affine transformations are applied during the training: rotation (± 5 degrees applied on one of the three axes), scaling (± 10% in the sagittal plane), and translation (± 3% in the sagittal plane).

## 2.4.2. Tumor segmentation

The tumor segmentation is multi-contrast since both gadolinium-enhanced T1w and T2w MRI scans carry valuable information. Being a multiclass task, the model was trained with a multiclass Dice loss, which is the average of Dice loss for each class. Patches of size 128x128x32 with a stride of 64x64x32 and a batch size of 8 were used to train the model. At inference time, the patches were stitched together by averaging for voxels having more than one prediction value. The same parameters used for spinal cord localization for depth, early stopping, initial learning rate, and learning rate scheduler were selected. Multiclass segmentation of IMSCT is a more complex task than localizing the spinal cord and required a higher number of base filters to reach optimal performance, which was set to 16.

### 2.4.3. Dataset split

For both models, 60% of the subjects were used for training, 20% for validation, and 20% for testing. Since only 45 subjects had tumors located in the lumbar region, all subjects were included in the training set to maximize the exposition of the model to IMSCT located in the lower part of the spinal cord. Hence, the model has not been validated on lumbar tumors.

## 2.5. Postprocessing

Postprocessing steps are applied to the model's prediction. The steps include: binarization with a threshold of 0.5, filling holes, and removing tumor prediction smaller than 0.2 cm³, as well as edema and cavity predictions smaller than 0.5 cm³ to limit false positives and noise. These postprocessing steps can be customized when applying the model with SCT.

## 2.6. Evaluation

### 2.6.1. Segmentation pipeline

The Dice score was used to evaluate the segmentation performance of both spinal cord localization and tumor segmentation models. To assess the localization task, the inclusion of all the IMSCT volumes into their respective bounding box, generated by the first step of the pipeline, was verified. Other metrics used to evaluate the multiclass segmentation model were: tumor true positive detection rate, false negative detection rate, precision, recall, as well as absolute and relative volume difference. A structure was considered detected if there was an overlap of at least 6 mm³ between the ground truth and the prediction. This stands for true positive and negative detection rates. The relative volume difference was computed by subtracting the volume prediction to the volume of the ground truth and dividing by the ground truth volume. A negative relative volume difference signifies an over-segmentation from the model. The absolute volume difference is the absolute value of the relative volume difference. 12 cross-validations were performed on the dataset in order to derive meaningful statistics on the results.

### 2.6.2. Comparison between cascaded architecture and single-step architecture

Training and inference time as well as Dice scores of the cascaded architecture were compared to the pipeline without the first step of the model. The same patch size and stride were used for both approaches. Since a single epoch takes several hours for the segmentation model without the initial cropping step and dozens of epochs are needed, only one training was performed without the first step of the pipeline (i.e., automatic cropping around the spinal cord). To ensure comparability of the two models, the same dataset split was used for both models. The inference time was computed on 100 images with average dimensions of 545x545x20 on a regular computer (i.e., without GPU) with 16 GB of RAM.

## 2.7. Implementation

The implementation of the models and pipeline is publicly available and can be found in the deep learning toolbox ivadomed v2.6.1 (http://ivadomed.org/). The packaged and ready-to-use model can be applied to custom data via SCT's function sct_deepseg[1]. The models were converted into the ONNX (Open Neural Network Exchange) format for fast prediction on CPUs.

# 3. Results

## 3.1. Spinal cord localization model

The spinal cord localization model reached a Dice score of 88.7% on the test set. When verifying if all tumor segmentations were contained in the bounding box generated by the first step of the pipeline, only two subjects (i.e., success rate of 99.4%) had part of their cavity located out of the bounding box. For both patients, the cavity rose up to the level of the brainstem, which explains why the model did not include this section (cavity extending out of the spinal cord region). Despite this partial truncation, the model correctly located the spinal cord area. Hence, the model successfully located all spine regions for all subjects.

## 3.2. Tumor segmentation

Figure 3 illustrates qualitative results from four testing subjects. The first row represents a subject where the segmentation for every class was successful (Dice > 85% for all structures). The second and third rows illustrate two examples of false positive detection for cavity and edema, respectively. A hyperintense signal on the T2w scan (white arrows), similar to what can be seen for liquid-filled cavities, led to the false identification of cavity tissue. The false positive edema classification was caused by a moderate hyperintense signal on the T2w image (yellow arrows) that can be confused with edema. The last row shows an example where the model failed to correctly identify the different tumor's structure. As seen in the third row, a moderate hyperintense signal on the T2w image generated a false positive edema prediction (pink arrows). The tumor was also misclassified, probably due to a strong hyperintense signal on T2w combined with a dim enhancement on the gadolinium-enhanced T1w (blue arrows). Cavities usually have a hyperintense signal with no enhancement on gadolinium-enhanced T1w.

---

[1] https://spinalcordtoolbox.com/en/stable/user_section/command-line.html#sct-deepseg

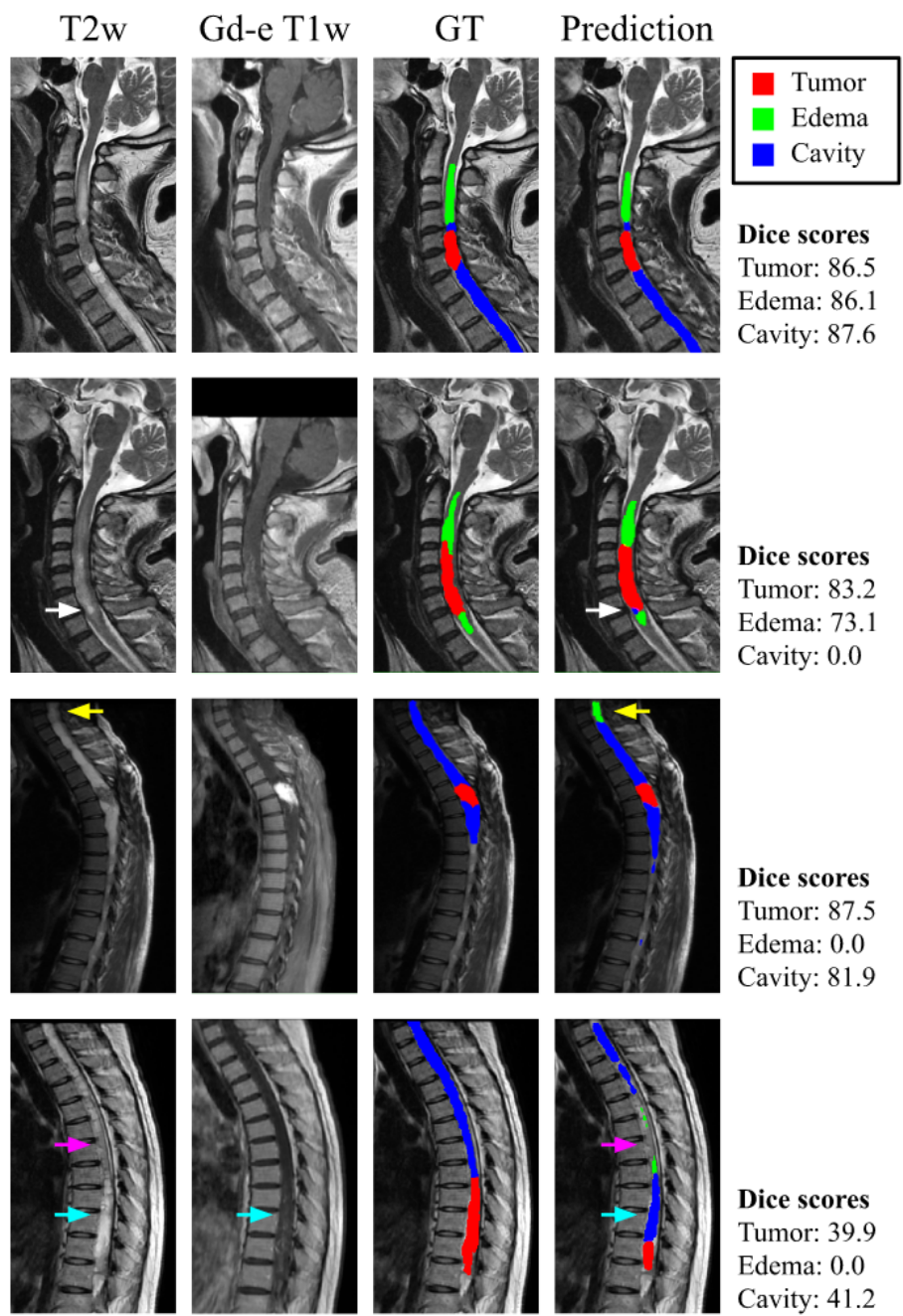

Figure 3: Sagittal MRIs showing the model's prediction on a subject containing tumor, edema, and cavity. Column 1 presents T2w input image, column 2 Gadolinium-enhanced T1w input image, column 3 the manual annotation, and column 4 the model's prediction. The white arrows point to a false positive cavity segmentation due to a hyperintense signal on T2w. The yellow arrows show a false positive edema segmentation, a consequence of a moderate hyperintense signal on T2w. The pink arrows point towards a false negative cavity and false positive edema segmentations caused by a moderate hyperintense signal on the T2w scan. The blue arrows indicate a misclassification of tumor tissue due to a hyperintense signal on T2w, also characteristic of liquid-filled cavities, and a moderate enhancement on the Gd-e T1w scan. Abbreviations: Gd-e: Gadolinium-enhanced, GT: Ground truth.

Table 3 presents the principal metrics for IMSCT related structures. An average Dice score across all cross-validations for tumor, cavity, and edema as a single class reached 76.7 ± 1.5%. When identifying each structure separately, the Dice score drops to 61.8%, 55.4%, and 31.0% for tumor, cavity, and edema, respectively. The lower scores associated with cavity and edema are partly due to false positive detection of the structure (e.g., the false detection rate of 46.3% for edema) as presented in the second and third rows of Figure 3. False positive detections of structure absent in the subject are systematically associated with a Dice score of 0% even if the predicted volume is a few voxels. When keeping only the subjects with edema or cavity, the Dice scores improve to 47.4% for edema and 65.2% for cavities (i.e., increase of respectively 16.4% and 9.8%). However, there is a relatively high true positive detection rate for all structures, i.e., higher than 87% for each class, meaning that when a structure is present, the model usually detects it. As for the volume difference, when segmenting all structures as a single class, the relative volume difference is -3.4%, and the absolute volume difference is 22.1%. The volume difference has also a performance drop when analyzing the structure separately. The model has a tendency to over-segment, i.e., negative relative volume difference, except for the edema, which is slightly under-segmented.

Table 3: Multiclass IMSCT segmentation metrics on 12 cross-validations (MEAN ± STD).

| Labels | Tumor + Cavity + Edema | Tumor | Cavity | Edema |
|---|---|---|---|---|
| Dice score [%] *Optimal value: 100* | 76.7 ± 1.5 | 61.8 ± 4.0 | 55.4 ± 4.4 | 31.0 ± 4.6 |
| True positive detection rate [%] *Optimal value: 100* | 98.7 ± 0.9 | 90.6 ± 4.1 | 93.8 ± 2.9 | 87.3 ± 6.4 |
| False positive detection rate [%] *Optimal value: 0* | 9.9 ± 3.5 | 17.2 ± 5.1 | 23.3 ± 4.3 | 46.3 ±7.2 |
| Precision [%] *Optimal value: 100* | 78.7 ± 2.6 | 74.3 ± 4.4 | 60.5 ± 3.7 | 41.4 ± 5.8 |
| Recall [%] *Optimal value: 100* | 77.5 ± 2.7 | 64.3 ± 4.0 | 67.5 ± 3.8 | 46.9 ± 7.5 |
| Absolute volume difference [%] *Optimal value: 0* | 22.1 ± 6.0 | 75.3 ± 39.5 | 61.8 ± 12.2 | 67.0 ± 31.7 |
| Relative volume difference [%] *Optimal value: 0* | -3.4 ± 9.9 | -26.0 ± 45.9 | -34.4 ± 33.3 | 4.5 ± 27.6 |

## 3.3. Advantages of a cascaded architecture

The cascaded architecture is associated with faster training time, inference time, and segmentation performance compared with the pipeline without the initial cropping step.

### 3.3.1. Training and inference time

Training a single epoch using the same patch and batch sizes without the first step of the cascaded architecture took ~3.4 hours, while one epoch lasted ~14 minutes for the proposed cascaded pipeline on a single NVIDIA Tesla P100 GPU.

The average inference time for the cascaded architecture took 22 seconds per image. In comparison, the model without the cropping step took 48 seconds. These times include the loading, resampling, preprocessing, spinal cord localization, multiclass tumor segmentation, postprocessing, and saving of the images. This time difference can become considerable when the model is sequentially applied to a large number of images.

### 3.3.2. Segmentation performance

In comparison with the single model architecture, the model using a cascaded architecture yielded a Dice score improvement of 5.0%, 30.0%, and 4.6% for the tumor, edema, and cavity respectively. The important class imbalance between the tumor and the background class paired with an imbalance of the edema class in the whole dataset led to consistent empty predictions for the edema class by the model using only the second step of the pipeline.

# 4. Discussion

We introduced an IMSCT segmentation model publicly available that can be applied to custom images within seconds with SCT. The cascaded architecture of the pipeline mitigates the class imbalance, speeds up the training and inference, and is associated with higher segmentation performance. The first step of the pipeline localizes the spinal cord and crops the image to isolate the region of interest. The robustness of the detection step is crucial since the segmentation model depends on this task (Gros et al., 2019). A Dice score of 88.7% on the test set was reached for this step and the spinal cord was correctly located for all subjects. The segmentation of the tumor, edema, and cavity as a single class reached 76.7% of Dice score. Segmenting the tumor, cavity, and edema separately is more challenging for the model yielding Dice scores of 61.8%, 55.4%, and 31.0%, respectively. The performance is hampered by false positive detections due to class imbalance throughout the dataset (not all subjects have a cavity or edema). However, the model has a high true positive detection rate (>87%) for every class.

## 4.1. Tissue heterogeneity

As seen in Table 3, the model reaches good performance when segmenting all structures associated with IMSCT together, but the metrics drop when segmenting each structure separately. The varying intensity patterns and the ambiguous delimitation of each structure can partly explain this behavior. On T2w scans, tumors can present hyperintense signals which can be confused with the hyperintense signal associated with liquid-filled cavities. A hyperintense signal on gadolinium-enhanced T1w modality helps to distinguish a tumor from a cavity, but tumors don't always exhibit an enhancement (see Figure 1A and 1D). The varying intensity patterns for tumors cause misclassification. The same goes for distinguishing edema from tumors or cavities from edema. For example, a cavity presenting a low voxel intensity can be mistaken for edema (see Figure 3 third row). Another challenging aspect of differentiating each component of IMSCT is the ill-defined boundaries between each structure. Figure 4 illustrates this phenomenon. The tumor displays edema intensity patterns on the T2w scan and has poor enhancement on gadolinium-enhanced T1w scan. The boundaries of the tumor are ambiguous which is challenging for clinicians, hence for the deep learning model.

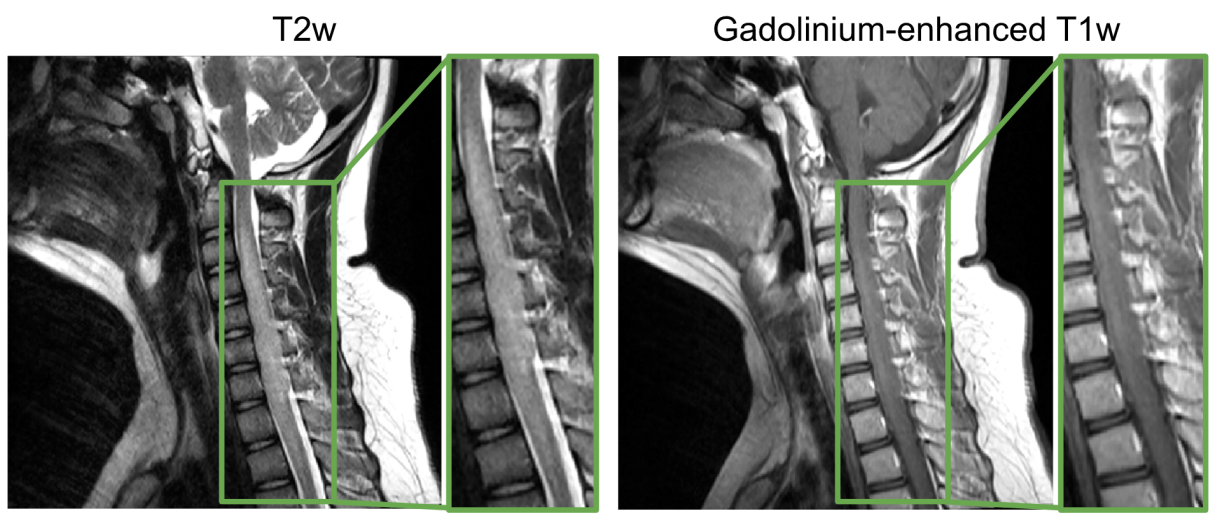

Figure 4: Subject with an astrocytoma with unclear tumor boundaries.

## 4.2. Class imbalance for edema and cavity

Another challenge that explains the lower performances for separate segmentation of tumor, edema, and cavity is the class imbalance of edema and cavity across the dataset. 67% of subjects have a cavity and only 51% of subjects have edema. The lower number of examples presented combined with the small size of edema and/or cavity compared to the background class (i.e., background represents 99.5% of voxels) hinder the model's ability to properly generalize leading to a high number of false positives. False positive detections of a structure are associated with a Dice score of 0% which highly impacts the metrics. This false positive rate is coupled with a high true positive rate meaning that when a structure is present the model usually detects it which is generally prefered in clinical settings.

## 4.3. Usage in clinical settings

The medical field can benefit from the integration of deep learning models into clinical settings (Perone and Cohen-Adad, 2019; Valliani et al., 2019). Even if imperfect, segmentation models can save hours of work to clinicians requiring only modifications to the automatic prediction. Also, the automated segmentation opens the door to investigate the clinical significance of spinal cord tumor characteristics (location, size) including assessing clinical disability (sensorimotor deficits), correlating to molecular status (e.g., H3 K27M mutation), evaluation of treatment effect, and predicting the prognosis. For instance, the work of Eden et al. quantified the impact of multiple sclerosis lesion distribution across various spinal white matter tracts (Eden et al., 2019), using a spinal atlas-based analysis (Lévy et al., 2015). Similar investigations could be done with spinal cord tumors.

## 4.4. Perspectives

### 4.4.1. Continuous learning

Continuous learning is increasingly popular in the medical field (Gonzalez et al., 2020; Pianykh et al., 2020) since it facilitates multiple center data training while addressing privacy issues related to data sharing (Gonzalez et al., 2020). Integrated tools for continual training would benefit the IMSCT model. Continuous learning allows the model to learn from new examples presented to the model, and become more and more robust through time. A challenge associated with continual training is catastrophic forgetting which occurs when the model gives disproportionate weight to new learning examples, hence forgetting the initial training data (Gonzalez et al., 2020; Kirkpatrick et al., 2017). Implementing techniques properly addressing this issue would allow to develop a robust continuous learning pipeline. This opens the possibility of having a cooperative model trained on data from different centers without direct data sharing that would improve through time.

### 4.4.2. Missing modalities

In this work, a multi-contrast model was chosen to improve the robustness of the segmentation. The presented model requires both T2w and gadolinium-enhanced T1w contrasts to generate a prediction. Users missing one of the contrasts would be unable to use the model, but solutions exist to mitigate this problem. HeMIS (Havaei et al., 2016) is a deep learning approach addressing this issue. A model robust to missing modalities would benefit the users since less images are necessary to have a prediction. Future work could focus on a HeMIS U-Net version of the model presented here.

### 4.4.3. SoftSeg

Recent work praises the benefits of leaning towards a soft training approach (Gros et al., 2020a; Müller et al., 2019). SoftSeg (Gros et al., 2020a) is a training pipeline yielding and propagating soft values to address partial volume effect and certainty calibration of the model for segmentation tasks. Gros et al. reports higher Dice scores when using SoftSeg compared to a conventional segmentation pipeline. SoftSeg could be generalized to multiclass predictions and be applied to IMSCT segmentation. This method takes into account partial volume effect, hence could benefit the volume measurement of unhealthy tissues. Also, the soft predictions would give more insight on the model's certainty especially towards ill-defined boundaries, which should be associated with higher uncertainty. The broader range of prediction values yields more information and allows a more enlightened postprocessing.

# 5. Conclusion

In this work, we presented the first model for multiclass segmentation for IMSCT with a two-stage U-Net based cascaded architecture. The choice of a cascaded architecture is associated with faster training and inference time, as well as higher dice scores. An average Dice score across 12 cross-validations of 76.7% was reached to segment all IMSCT structures as a single class. Varying intensity patterns, ill-defined boundaries, and high class imbalance was a challenge to the separate labeling of tumors, cavities, and edemas. The model is associated with true positive detection rates above 87% for all components (tumor, edema, and cavity). Future work could focus on the implementation of continual learning tools, techniques to address missing modalities, or a SoftSeg version of the model. The segmentation pipeline is available in the SCT software and can be applied to custom data through a single command-line in a few seconds.

# 6. Acknowledgements

The authors want to thank Olivier Vincent, Lucas Rouhier, Anthime Bucquet, Joseph Paul Cohen, and Sara Paperi for insightful discussion during the development of the model.

Funding: This work was supported by IVADO [EX-2018-4], Canada Research Chair in Quantitative Magnetic Resonance Imaging [950-230815], the Canadian Institute of Health Research [CIHR FDN-143263], the Canada Foundation for Innovation [32454, 34824], the Fonds de Recherche du Québec - Santé [28826], the Fonds de Recherche du Québec - Nature et Technologies [2015-PR-182754], the Natural Sciences and Engineering Research Council of Canada [RGPIN-2019-07244], the Canada First Research Excellence Fund (IVADO and TransMedTech), the Courtois NeuroMod project and the Quebec BioImaging Network [5886, 35450], Spinal Research and Wings for Life (INSPIRED project), the National Science Foundation of China [81870958, 81571631], the Beijing Municipal Natural Science Foundation for Distinguished Young Scholars [JQ20035], and the Special Fund of the Pediatric Medical Coordinated Development Center of Beijing Hospitals Authority [XTYB201831]. A.L. has a fellowship from NSERC and FRQNT. The authors thank the NVIDIA Corporation for the donation of a Titan X GPU.